\documentclass[12pt]{article}
\usepackage{epsfig}
\usepackage{amsfonts}
\begin{document}
\begin{titlepage}
\begin{center}

{\Large Lagrangian acceleration statistics in turbulent flows}

\vspace{2.cm}

{\bf Christian Beck}

\vspace{0.5cm}

School of Mathematical Sciences, Queen Mary, University of London,
Mile End Road, London E1 4NS.

\end{center}

\vspace{4cm}

\abstract{We show that the probability densities of accelerations
of Lagrangian test particles in turbulent flows as measured by
Bodenschatz et al. [Nature 409, 1017 (2001)] are in excellent
agreement with the predictions of a stochastic model introduced in
[C. Beck, PRL 87, 180601 (2001)] if the fluctuating friction
parameter is assumed to be log-normally distributed. In a
generalized statistical mechanics setting, this corresponds to a
superstatistics of log-normal type. We analytically evaluate all
hyperflatness factors for this model and obtain a flatness
prediction in good agreement with the experimental data. There is 
also good agreement with DNS data of Gotoh et al.
We relate the model to a generalized
Sawford model with fluctuating parameters, and discuss a possible
universality of the small-scale statistics.}

\vspace{1.3cm}

\end{titlepage}

In the past two years there has been considerable experimental
progress in mesasuring the statistics of Lagrangian test particles
in turbulent flows \cite{boden,pinton,boden2}. While simple
dynamical models like the Sawford model \cite{saw,pope} predict a
Gaussian acceleration statistics, the measurements of Bodenschatz
et al.\ \cite{boden, boden2} have confirmed that the probability
distribution of acceleration $a$ is strongly non-Gaussian and has
strongly pronounced tails. As shown in \cite{pla,prl}, the
measured acceleration distributions are well fitted by Tsallis
distributions \cite{tsa1} for $|a|<30$, in units of the standard
deviation. However, very new recent measurements \cite{boden3} and
numerical simulations \cite{gotoh} indicate that there are
deviations from Tsallis statistics for extremely large
accelerations $|a|>30$, i.e.\ for extremely rare events of very strong
forces acting on the particle.

In this Letter we show
that the simple dynamical model
previously introduced in \cite{prl} can be generalized in such a
way that it yields quite a perfect description of the observed
Lagrangian acceleration statistics in turbulent flows. The
coincidence is better than in previous models based on Tsallis
statistics. In the language of statistical mechanics, the relevant
new statistics that fits the data quite perfectly is
`superstatistics of log-normal type'. While it can be proved that
log-normal superstatistics, as any superstatistics, is close to
Tsallis statistics for not too large accelerations, there are
significant differences in the tails.

The concept of superstatistics was introduced in \cite{eddie}. A
superstatistics is a 'statistics of a statistics', one given by
ordinary Boltzmann factors $e^{-\beta E}$ and the other one by the
distribution function of $\beta$, which is assumed to be
fluctuating. Superstatistics are generally relevant for an
effective description of driven nonequilibrium systems with
stationary states which possess a fluctuating intensive parameter
$\beta$. This can, for example, be the inverse temperature or, in
the turbulence application, the energy dissipation rate. Tsallis
statistics is a special superstatistics obtained for a
$\chi^2$-distribution of the intensive parameter. But
as shown by Tsallis and Souza \cite{tsa2}, more general versions
of statistical mechanics can also be constructed for other
superstatistics than Tsallis statistics, in particular for the
superstatistics of log-normal type which turns out to be relevant
here.

A superstatistics can be easily dynamically realized, by
generalizing the approach of \cite{prl}. Consider a Langevin of
the form
\begin{equation}
\dot{a}= \gamma F(a) +\sigma L(t), \label{1}
\end{equation}
where $L(t)$ is Gaussian white noise, $\gamma >0$ is a friction
constant, $\sigma$ describes the strength of the noise, and $F(a)=
- \frac{\partial}{\partial a}V(a)$ is a drift force. For most
applications, it is sufficient to consider linear forces $F(a)=-a$
as generated by $V(a)=\frac{1}{2}a^2$. For an ordinary Brownian
particle, $a$ would be the velocity, but for the Lagrangian
turbulence application, $a$ actually stands for the acceleration
of the Lagrangian test particle, i.e.\ a velocity difference on a
very small time scale. We will consider linear Langevin equations
of the form (\ref{1}) for each component of the acceleration.

If $\gamma$ and $\sigma$ are constant then the stationary
probability density of $a$ is  $p(a)\sim e^{-\beta V(a)}$, where
$\beta:=\frac{\gamma}{\sigma^2}$. However, since the energy
dissipation fluctuates in a turbulent flow we now let the
parameters $\gamma$ and $\sigma$ fluctuate such that $\beta
:=\frac{\gamma}{\sigma^2}$ has some probability density
$f(\beta)$. These parameter fluctuations are assumed to be on a
relatively long time scale (or a relatively large spatial scale)
so that the system can temporarily reach local equilibrium.
The Lagrangian test particle moves through these different regions
with different $\beta$.

One
obtains for the conditional probability $p(a|\beta)$ (i.e. the
probability of $a$ given some value of $\beta$)
\begin{equation}
p(a|\beta)=\frac{1}{Z(\beta)}\exp \left\{ -\beta V(a) \right\},
\end{equation}
where $Z(\beta )$ is a normalization constant, and for the joint
probability $p(a,\beta)$ (i.e. the probability to observe both a
certain value of $a$ and a certain value of $\beta$)
\begin{equation}
p(a,\beta)=p(a|\beta)f(\beta),
\end{equation}
and finally for the marginal probability $p(a)$ (i.e.\ the
probability to observe a certain value of $a$ no matter what
$\beta$ is)
\begin{equation}
p(a)=\int p(a|\beta)f(\beta)d\beta . \label{9}
\end{equation}
In \cite{prl} a $\chi^2$-distribution was chosen for $f(\beta)$,
and it was shown that this generates Tsallis statistics for the
marginal distribution, i.e.\ $p(a)\sim (1+ \tilde{\beta}(q-1)
V(a))^{-1/(q-1)}$, where $q$ is the entropic index of nonextensive
statistical mechanics \cite{tsa1}.

Let us here choose another distribution for $f(\beta)$, in fact  one
of the examples dealt with in \cite{eddie}, the log-normal
distribution
\begin{equation}
f(\beta) = \frac{1}{\beta s \sqrt{2\pi}}\exp\left\{ \frac{-(\log
\frac{\beta}{m})^2}{2s^2}\right\}, \label{Logno}
\end{equation}
where $m$ and $s$ are parameters. The average $\beta_0$ of the
above log-normal distribution is given by $\beta_0=m\sqrt{w}$ and
the variance by $\sigma^2=m^2w(w-1)$, where $w:= e^{s^2}$. For our
Lagrangian turbulence model it is sufficient to consider linear
drift forces $F(a)=-a$. The integral given by (\ref{9})
\begin{equation}
p(a) = \frac{1}{2\pi s }\int_0^\infty d\beta \; \beta^{-1/2}
\exp\left\{ \frac{-(\log \frac{\beta}{m})^2}{2s^2}\right\}
e^{-\frac{1}{2}\beta a^2} , \label{10}
\end{equation}
cannot be evaluated in closed form, but the equation is easily
numerically integrated, and can be compared with experimentally
measured densities $p(a)$. Similar distributions as in (\ref{10})
have also been considered in \cite{cast}, however without a
dynamical realization in terms of a stochastic differential
equation. The distribution $p(a)$ has variance 1 for the choice
$m=\sqrt{w}$, hence only one fitting parameter $s$ remains if one
compares with experimental data sets that have variance 1.

Fig.~1 shows the histograms of accelerations of Lagrangian test
particles as measured by Bodenschatz et al. \cite{boden,boden2}.
The distributions are rescaled to variance 1. As shown
previously \cite{pla,prl}, the experimental distributions are
reasonably well approximated by Tsallis distributions for $|a|<30$.
But log-normal superstatistics yields a better fit, using
only one fitting parameter $s$. The measured distributions for
$R_\lambda =690$ and $R_\lambda = 970$ are basically the same and
very well fitted by eq.~(\ref{10}) with $s^2=3.0$. Note that $m$
is not a free fitting parameter but fixed as
$m=\sqrt{w}=e^{\frac{s^2}{2}}$ to give variance 1. Since
Bodenschatz's data reach rather large accelerations $a$ (in units
of the standard deviation), the measured tails of the
distributions allow for proper distinction between various
superstatistics. The main difference between the
$\chi^2$-superstatistics studied in \cite{prl} and log-normal
superstatistics studied here is the fact that $p(a)$ decays with a
power law for the former ones, whereas it decays with a more
complicated logarithmic decay for the latter ones. For alternative
fitting attempts based on multifractal models, see \cite{ari}.

All moments of the distribution (\ref{10}) exist and can be easily evaluated
as follows. The moments of a Gaussian distribution
\begin{equation}
p_G(a)=\sqrt{\frac{\beta}{2\pi}}e^{-\frac{1}{2}\beta a^2}
\end{equation}
are given by
\begin{equation}
\langle a^r \rangle_G=\frac{1}{\beta^{r/2}} (r-1)!! \label{gauss}
\end{equation} ($r$ even). Moreover, the moments of the lognormal
distribution (\ref{Logno}) are given by
\begin{equation}
\langle \beta^r \rangle_{LN} = m^r w^{\frac{1}{2}r^2}.
\label{logno}
\end{equation}
Combining eq.~(\ref{gauss}) and (\ref{logno}) one obtains the
moments of the distribution $p(a)$ given by eq.~(\ref{10}) as
\begin{eqnarray}
\langle a^r \rangle &=&\langle \langle a^r\rangle_G\rangle_{LN}
\\ &=& (r-1)!! \langle \beta^{-r/2}\rangle_{LN}
\\
&=&(r-1)!! m^{-\frac{r}{2}} w^{\frac{1}{8}r^2}
\end{eqnarray}
In particular, the variance is given by
\begin{equation}
\langle a^2 \rangle = m^{-1} \sqrt{w}.
\end{equation}
All hyperflatness factors $F_r$ are independendent of $m$ and
given by
\begin{equation}
F_r:=\frac{\langle a^{2r} \rangle}{\langle a^2 \rangle^r}=
(2r-1)!! w^{\frac{1}{2}(r-1)}.
\end{equation}
In particular, the flatness $F_2$ is given by
\begin{equation}
F_2:=\frac{\langle a^4\rangle}{\langle a^2\rangle}=3w=3e^{s^2}.
\label{flat}
\end{equation}
Measuring the flatness $F_2$ of some experimental data thus
provides a very simple method to determine the fitting parameter
$s$ of lognormal superstatistics. Or, if $s$ is fitted from the
shape of the densities, then a prediction on the flatness factor
can be given. Note that $w$ is the analogue of $q$ in the
nonextensive approach \cite{eddie}.

Using the parameter $s^2=3.0$ that yields a good fit of the
densities measured by Bodenschatz at $R_\lambda =690$ and
$R_\lambda =970$, a flatness of $F_2=60.3$ is predicted from
eq.~(\ref{flat}), in agreement with the measured flatness values
reported in \cite{boden2,boden3}.

In direct numerical simulations (DNS) of the Navier-Stokes
equations, even larger accelerations $a$ can be reached. Fig.~2
shows Gotoh's results on the pressure distribution as obtained by
DNS at $R_\lambda=380$ \cite{gotoh}. In good approximation the pressure
statistics coincides with the acceleration statistics of a
Lagrangian test particle. Gotoh's histograms reach accelerations
up to 150 (in units of the standard deviation), a much larger
statistics than can be presently reached in Bodenschatz's
experiment. Hence the tails of these distributions can very
sensitively distinguish between various superstatistics models.
Fig.~2 shows that log-normal superstatistics with $s^2=3.0$ again
yields a good fit, keeping in mind that one compares data that
vary over 12 orders of magnitude. The fit quality can be slightly
further improved if one uses for $f(\beta)$ a log-normal
distribution that has a lower and upper cutoff
($\beta_{min}\approx 0.0005$ and $\beta_{max}\approx 25$). The
above truncation may effectively model finite Reynolds number and
finite size effects, which are certainly present in any numerical
simulation of the Navier-Stokes equation.

If the Reynolds number is sufficiently large ($R_\lambda > 300$),
we notice that the same fitting parameter $s^2=3.0$ yields a good
fit of all 3 different data sets ($R_\lambda =380, 690, 970$). It
seems that $s^2$ is essentially independent of the Reynolds
number for large $R_\lambda$, 
in agreement with the measured approximate Reynolds number
independence of the flatness reported in \cite{boden2} for
$R_\lambda > 300$. If this tendency is confirmed by further
experiments then it seems natural to conjecture that the variance
of the log-normal distribution converges to a finite universal
value for $R_\lambda \to \infty$. This value might simply be given
by $s^2=3$. The value 3 could stand in connection with the three
spatial degrees of freedom: As it is apparent from
eq.~(\ref{Logno}), the variable $Y:=\log \frac{\beta}{m}$ is a
Gaussian random variable with average 0 and variance $s^2$. The
fluctuating $\beta$ is related to the fluctuating energy
dissipation rate. Since energy can flow independently into the
three space directions, we may write
\begin{equation}
Y= \log \frac{\beta_x}{m} +\log \frac{\beta_y}{m} +\log
\frac{\beta_z}{m}, \label{Y}
\end{equation}
where each of the independent random variables $X_i:=\log
\frac{\beta_i}{m}$, $i=x,y,z$, is Gaussian with average 0. Our
fitting observation $s^2=3$ is thus equivalent to the fact that
each of the Gaussian random variables $X_i$, describing the energy
flow in the $i$-th direction, has variance 1. Whether this is a
random coincidence for finite Reynolds number
or whether this is a deep fundamental principle
underlying the small-scale statistics for $R_\lambda \to \infty$
is not clear at the moment.

What remains to be done is to precisely relate the parameter
$\beta$ of our superstatistics approach to the fluctuating energy
dissipation $\epsilon$ in the turbulent flow. For this we may just
compare with a previously studied Lagrangian model, the Sawford
model \cite{saw,pope,reynolds}. The Sawford model assumes that the
joint stochastic process $(a(t),u(t),x(t))$ of acceleration,
velocity and position of a Lagrangian test particle obeys the
stochastic differential equation
\begin{eqnarray}
\dot{a}& =&-(T_L^{-1}+t_\eta^{-1})a-T_L^{-1}t_\eta^{-1} u
\nonumber \\ &\,&
+\sqrt{2\sigma_u^2(T_L^{-1}+t_\eta^{-1})T_L^{-1}t_\eta^{-1}}\;
L(t)
\\ \dot{u} &=&a \\ \dot{x} &=&u,
\end{eqnarray}
where $L(t)$ is again Gaussian white noise. $T_L$ and $t_{\eta}$
are two time scales, with $T_L
>>t_\eta$. In this model one has
\begin{eqnarray}
T_L&=& \frac{2\sigma_u^2}{C_0 \bar{\epsilon}} \\
t_\eta &=& \frac{2a_0\nu^{1/2}}{C_0\bar{\epsilon}^{1/2}}, \\
\end{eqnarray}
where $\bar{\epsilon}$ is the average
energy dissipation, $C_0, a_0$ are Lagrangian structure function constants,
and $\sigma_u^2$ is the variance of the velocity distribution. The
Taylor scale Reynolds number is given by
\begin{equation}
R_\lambda = \frac{\sqrt{15}\sigma_u^2}{\sqrt{\nu \bar{\epsilon}}}.
\end{equation}
The  Sawford model predicts Gaussian stationary distributions for
$a$ and $u$, and is thus at variance with the recent measurements.
However, a straightforward idea is to generalize the Sawford model
with constant parameters to a generalized Sawford model with
fluctuating parameters, following a similar type of arguments as
in \cite{prl}. This was recently worked out by Reynolds
\cite{reynolds}, who emphasized compatibility of the obtained
stochastic model with Kolmogorov's refined similarity hypothesis.

We note that there are several possibilities to extend the Sawford
model with constant parameters to an extended one with fluctuating
ones, depending on which of the variables $\bar{\epsilon}$ in the
above model equations are replaced by a fluctuating $\epsilon$ and
which are not. For our purposes it is sufficient to consider the
Sawford model in the limit $T_L \to \infty$, which is a reasonable
approximation for large Reynolds numbers. In that limit the model
with constant parameters becomes identical to eq.~(\ref{1}) with
$F(a)=-a$ and with the identification
\begin{eqnarray}
\gamma &=&\frac{C_0}{2a_0} \nu^{-1/2} \bar{\epsilon}^{1/2}\label{gamma} \\
\sigma &=& \frac{C_0^{3/2}}{2a_0} \nu^{-1/2} \bar{\epsilon} \label{sigma}
\end{eqnarray}
This clarifies the physical meaning of the average of our
parameters $\gamma$ and $\sigma$. To proceed to a model with
fluctuting parameters, the simplest possibility is to replace in
both of the above equations $\bar{\epsilon}$ by a fluctuating
$\epsilon$. This yields
\begin{equation}
\beta = \frac{\gamma}{\sigma^2} = \frac{2a_0}{C_0^2} \nu^{1/2}
\epsilon^{-3/2}, \label{26}
\end{equation}
i.e. the fluctuating energy dissipation $\epsilon$ is proportional
to $\beta^{-2/3}$. Based on this equation, one can
derive a couple of interesting relations. For example, for the
constant $a_0$ defined by the relation
\begin{equation}
\langle a^2 \rangle =: a_0 \langle \epsilon \rangle^{3/2}
\nu^{-1/2}
\end{equation}
one obtains after a short calculation the relation
\begin{equation}
a_0 = \frac{1}{\sqrt{2}}C_0
e^{\frac{1}{12}s^2}
\end{equation}
Reynolds' DNS data suggest $C_0\approx 7$ \cite{reynolds}, hence
$s^2=3$ yields $a_0 \approx 6$, in agreement with Bodenschatz's
measurements \cite{boden2}. Moreover, for the Gaussian random
variable $Y=\log \frac{\beta}{m}$ studied in eq.~(\ref{Y}) one
obtains the relation
\begin{equation}
Y=-\frac{3}{2}\log \frac{\epsilon}{\langle \epsilon \rangle}
-\frac{1}{3}s^2
\end{equation}
independent of the value of $a_0$ and $C_0$.

However, in general there are also other possible dynamical
realizations of a Sawford model with
fluctuating parameters than eq.~(\ref{26}). For
example, we may keep $\gamma $ constant in eq.~(\ref{gamma}) and
let only $\sigma$ in eq.~(\ref{sigma}) fluctuate. This leads to
$\beta \sim \epsilon^{-2}$. Or, we may keep $\sigma$ constant and
let only $\gamma$ fluctuate. This leads to $\beta \sim
\epsilon^{1/2}$. All these cases have in common that $\beta \sim
\epsilon^\kappa$, where $\kappa$ is some power. Of course, for a
log-normal distribution the power $\kappa$ does not change the
functional form of the log-normal distribution, which once again
is a hint for the physical relevance of a log-normally distributed
$\beta$. While all the above possibilities lead to the same
marginal distribution $p(a)$ given by eq.~(\ref{10}), they can be
dynamically distinguished by looking at the acceleration
autocorrelation function, which exhibits different behaviour for
the various cases.


To summarize, we have shown that the measured densities in
Lagrangian turbulence experiments are very well described by
log-normal superstatistics. This superstatistics is easily
dynamically realized by considering a log-normal generalization of
the model previously studied in \cite{prl}. Only one parameter $s$
is fitted to obtain excellent agreement with the experimentally
measured distributions and the DNS data. Log-normal
superstatistics differs from $\chi^2$-superstatistics, i.e.\
ordinary Tsallis statistics, but for moderately large
accelerations Tsallis statistics is often a good approximation. A
possible hypothesis is that the variance parameter $s^2$ converges
for $R_\lambda \to \infty$ and is given by $s^2=3$. In that case
no free parameters would be left, and a universal Lagrangian small
scale statistics would arise.

\subsection*{Acknowledgement}
I am very grateful to Toshi Gotoh for providing me with the DNS
data displayed in Fig.~2.

\vspace{2cm}

\epsfig{file=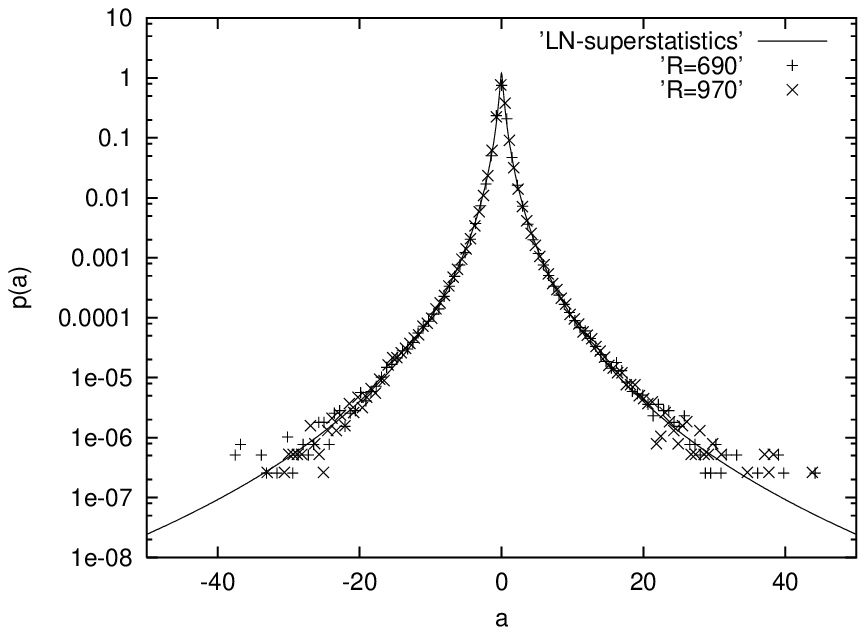}

\vspace{1cm}

{\bf Fig.1} Acceleration distributions as measured by Bodenschatz
et al. \cite{boden,boden2} for Reynolds number $R_\lambda =690$
and $970$, and the log-normal superstatistics distribution $p(a)$
given by eq.~(\ref{10}) with $s^2=3.0$.

\vspace{2cm}

\epsfig{file=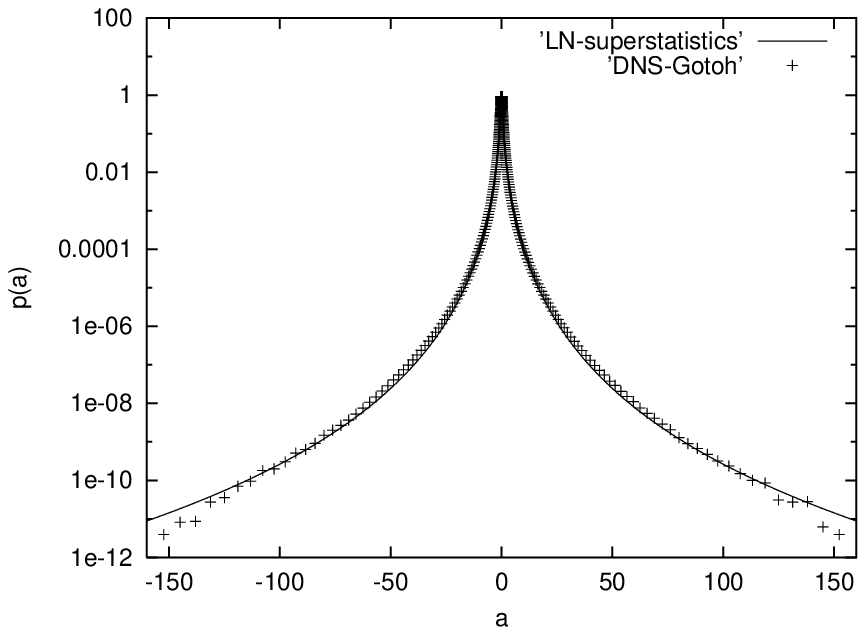}

\vspace{1cm}

{\bf Fig.2} DNS data obtained by Gotoh et al. \cite{gotoh}, and
the log-normal superstatistics distribution eq.~(\ref{10}) with
$s^2=3.0$.

\end{document}